\documentclass[aps,showpacs,superscriptaddress,preprint]{revtex4}%
\usepackage{amsfonts}
\usepackage{amsmath}
\usepackage{amssymb}
\usepackage{graphicx}%
\begin{document}
\title{Loschmidt Echo and Berry phase of the quantum system coupled to the $XY$ spin
chain: Proximity to quantum phase transition}
\author{Zi-Gang Yuan}
\affiliation{State Key Laboratory for Superlattices and Microstructures, Institute of
Semiconductors, Chinese Academy of Sciences, P.O. Box 912, Beijing 100083, China}
\author{Ping Zhang}
\affiliation{Institute of Applied Physics and Computational Mathematics, P.O. Box 8009,
Beijing 100088, China}
\author{Shu-Shen Li}
\affiliation{State Key Laboratory for Superlattices and Microstructures, Institute of
Semiconductors, Chinese Academy of Sciences, P.O. Box 912, Beijing 100083, China}
\keywords{Quantum phase transition, Loschmidt echo}
\pacs{75.10.Pq, 03.65.Vf, 05.30.Pr, 42.50.Vk}

\begin{abstract}
We study the Loschmidt echo (LE) of a coupled system consisting of a central
spin and its surrounding environment described by a general XY spin-chain
model. The quantum dynamics of the LE is shown to be remarkably influenced by
the quantum criticality of the spin chain. In particular, the decaying
behavior of the LE is found to be controlled by the anisotropy parameter of
the spin chain. Furthermore, we show that due to the coupling to the spin
chain, the ground-state Berry phase for the central spin becomes nonanalytical
and its derivative with respect to the magnetic parameter $\lambda$ in spin
chain diverges along the critical line $\lambda=1$, which suggests an
alternative measurement of the quantum criticality of the spin chain.

\end{abstract}
\maketitle

Quantum phase transition (QPT), which is closely associated with the
occurrence of nonanalyticity of the ground-state energy as a function of the
coupling parameters in the system's Hamiltonian\cite{Sach}, are of extensive
current interest, mainly in condensed matter physics because they are not only
at the origin of unusual finite temperature properties but also promote the
formation of new states of matter like unconventional superconductivity in
heavy-fermion system\cite{Mathur}. In the parameter space, the points of
nonanalyticity of the ground-state energy density are referred to as critical
points and define QPT. At these points one typically witnesses the divergence
of the length associated with the two-point correlation function of some
relevant quantum field. In experiments QPT has been extensively studied in the
heavy-fermion compounds\cite{Lace,Gege}. Recently, QPT has drawn a
considerable interest in other fields of physics. More specifically QPT has
been studied by analyzing scaling, asymptotical behavior and extremal points
of various entanglement measures\cite{Ost,Vidal,Chen,Gu,Wu}. The connection
between geometric Berry phase (BP) and QPT for the case of spin-XY model has
also been studied\cite{Car,Zhu,Ham}, through which a remarkable relation
between the BP and criticality of spin chains is established. In addition, a
characterization of QPT in terms of the overlap between two ground states
obtained for two different values of external parameters has been
presented\cite{Zarnidi}.

Another way to study quantum criticality is to investigate quantum dynamics of
the many-body systems. Recently, Sengupta \textit{et al}.\cite{Seng} have
studied time evolution of the Ising order correlations under a time-dependent
transverse field and shown that the order parameter is best enhanced in the
vicinity of the quantum critical point. Quan \textit{et al}.\cite{Quan} have
studied transition dynamics of a quantum two-level system from a pure state to
a mixed one induced by the quantum criticality of the surrounding many-body
system. They have shown that the decaying behavior of the LE is best enhanced
by the QPT of the surrounding system. Yi \textit{et al}.\cite{Yi} have
reported the relation between the Hahn spin echo of a spin-1/2 particle and
QPT in a spin chain which is coupled to the particle. It is expected that
further work associated with the dynamical measurement of QPT via a coupling
to the central probe system will be reported afterwards. From this aspect a
thorough theoretical investigation of the quantum dynamics in QPT regime,
including the various kinds of spin-chain models, is necessary and will be
helpful for future experimental references.

In this paper, we present a theoretical study of the behavior of the Loschmidt
echo (LE) of a coupled spin system which consists of two quantum subsystems.
One subsystem is characterized by a spin-1/2 Hamiltonian, which denotes the
general two-level particles. We call this subsystem the central spin, in the
sense that this spin plays the role of measuring apparatus. Whereas the other
subsystem plays the role of surrounding many-body environment and is modeled
by a general $XY$ spin chain in a transverse magnetic field. The present study
is directly motivated by the recent theoretical report\cite{Quan} that the
quantum critical behavior of environmental system strongly affects its
capability of enhancing the decay of LE. Here we extend the Ising model used
in Ref.[15] for simulating the environmental subsystem to the more general XY
model. Compared to the Ising model, the XY model is parametrized by $\gamma$
and $\lambda$ [see Eq. 1(b) below]. Two distinct critical regions appear in
parameter space: the segment $(\gamma,\lambda)=(0,(0,1))$ for the $XX$ spin
chain and the critical line $\lambda_{c}=1$ for the whole family of the $XY$
model\cite{Sach}. The behavior of decaying enhancement of the LE calculated in
Ref.[15] can be used as a measure of the presence of the quantum criticality
of the Ising spin chain. It remains yet to be exploited whether this decaying
enhacement sustains in the whole critical regions for the $XY$ model.

The other interest in this paper is to study the BP properties of the coupled
system. Instead of investigating the BP of the environmental $XY$ spin chain
which has been previously studied\cite{Car,Zhu,Ham}, we focus our attention to
the ground-state BP of the central quantum subsystem. Due to the coupling, it
is expected that the quantum criticality of the surrounding $XY$ spin chain
will influence the BP of the central spin, which is found in this paper to be
close proximity to the nonanalytical and divergent behavior of QPT of the
environmental spin chain in the critical region.

We consider a two-level quantum system (central spin) transversely coupled to
a environmental spin chain which is described by the one-dimensional $XY$
model. The corresponding Hamitonian is given by $H=H_{C}+H_{E}+H_{I}$, where
(we take $\hbar=1$)%
\begin{equation}
H_{C}=\mu\sigma^{z}/2+\nu\sigma^{x}/2, \tag{1a}%
\end{equation}%
\begin{equation}
H_{E}=-\sum_{l}^{N}\left(  \frac{1+\gamma}{2}\sigma_{l}^{x}\sigma_{l+1}%
^{x}+\frac{1-\gamma}{2}\sigma_{l}^{y}\sigma_{l+1}^{y}+\lambda\sigma_{l}%
^{z}\right)  , \tag{1b}%
\end{equation}%
\begin{equation}
H_{I}=\frac{g}{N}\overset{N}{\underset{l=1}{\sum}}\sigma^{z}\sigma_{l}^{z}.
\tag{1c}%
\end{equation}
Here the Pauli matrices $\sigma^{\alpha}$ ($\alpha=x,y,z$) and $\sigma
_{l}^{\alpha}$ are used to describe the central spin and the environmental
spin-chain subsystems, respectively. The parameter $\lambda$ in $H_{E}$ is the
intensity of the magnetic filed applied along $z$-axis, and $\gamma$ measures
the anisotropy in the in-plane interaction. It is well known that the $XY$
model in Eq. (1b) encompasses two other well-known spin models: it turns into
transverse Ising chain for $\gamma=1$ and the $XX$ chain for $\gamma=0$.
$H_{I}$ gives the coupling between the central spin and the surrounding spin
chain. The above employed model is similar to the Hepp-Coleman
model\cite{Hepp,Bell} or its generalization\cite{Naka,Cini,Sun}.

As for quantum criticality in the $XY$ model, there are two universality
classes depending on the anisotropy $\gamma$. The critical features are
characterized in terms of a critical exponent $\nu$ defined by $\xi
\sim|\lambda-\lambda_{c}|^{-\nu}$ with $\xi$ representing the correlation
length. For any value of $\gamma$, quantum criticality occurs at a critical
magnetic field $\lambda_{c}=1$. For the interval $0<\gamma\leq1$ the model
belongs to the Ising universality class characterized by the critical exponent
$\nu=1$, while for $\gamma=0$ the model belongs to the $XX$ universality class
with $\nu=1/2$\cite{Sach}.

Following Ref.[15], we assume that the central spin is initially in a
superposition state $|\phi_{S}(0)\rangle=c_{g}|g\rangle+c_{e}|e\rangle$, where
$|g\rangle=\left(  \sin\frac{\theta}{2},-\cos\frac{\theta}{2}\right)
^{\text{T}}$ and $|e\rangle=\left(  \cos\frac{\theta}{2},\sin\frac{\theta}%
{2}\right)  ^{\text{T}}$ with $\theta=\tan^{-1}(\nu/\mu)$ are ground and
excited states of $H_{C}$, respectively. The coefficients $c_{g}$ and $c_{e}$
satisfy the normalization condition, $|c_{g}|^{2}+|c_{e}|^{2}=1$. Then the
evolution of the $XY$ spin chain initially prepared in $|\varphi(0)\rangle$,
will split into two branches $|\varphi_{\alpha}(t)\rangle=\exp(-iH_{\alpha
}t)|\varphi(0)\rangle$ ($\alpha=g,e$), and the total wave function is obtained
as $|\psi(t)\rangle=c_{g}|g\rangle\otimes|\varphi_{g}(t)\rangle+c_{e}%
|e\rangle\otimes|\varphi_{e}(t)\rangle$. Here, the evolutions of the two
branch wave functions $|\varphi_{\alpha}(t)\rangle$ are driven, respectively,
by the two effective Hamiltonians
\begin{equation}
H_{g}=\langle g|H|g\rangle=H_{E}-\delta\underset{l=1}{\overset{N}{%
{\displaystyle\sum}
}}\sigma_{l}^{z}-\Delta, \tag{2a}%
\end{equation}%
\begin{equation}
H_{e}=\langle e|H|e\rangle=H_{E}+\delta\underset{l=1}{\overset{N}{%
{\displaystyle\sum}
}}\sigma_{l}^{z}+\Delta, \tag{2b}%
\end{equation}
where $\Delta=\sqrt{\mu^{2}+\nu^{2}}/2$ and $\delta=g\cos\theta/N$. Obviously,
both $H_{g}$ and $H_{e}$ describe the $XY$ model in a transverse field, but
with a tiny difference in the field strength. The central spin in two
different states $|g\rangle$ and $|e\rangle$ will exert slightly different
backactions on the surrounding spin chain, which manifests as two effective
potentials $V_{g}=-\delta\underset{l=1}{\overset{N}{%
{\displaystyle\sum}
}}\sigma_{l}^{z}$ and $V_{e}=\delta\underset{l=1}{\overset{N}{%
{\displaystyle\sum}
}}\sigma_{l}^{z}$. This difference results in the decay of the LE\cite{Kark}
defined as\cite{Quan}
\begin{equation}
L(t)=|\langle\varphi_{g}(t)|\varphi_{e}(t)\rangle|^{2} \tag{3}\label{e3}%
\end{equation}
The LE has been proved to be conveniently related to depicting quantum
decoherence of the central system\cite{Quan}: consider the purity
defined\cite{Kark} by $P=$Tr$_{C}(\rho_{C}^{2})=$Tr$_{C}\{[$Tr$_{E}%
\rho(t)]^{2}\}$. Here $\rho(t)=|\psi(t)\rangle\langle\psi(t)|$, and
Tr$_{C(E)}$ means tracing over the degrees of freedom for the central spin
(environmental spin chain). A straightforward calculation reveals the
relationship between the LE and the purity as $P=1-2|c_{g}c_{e}|^{2}\left[
1-L(t)\right]  $\cite{Quan}. This equation indicates that the purity depends
on the initial state of the central spin and the surrounding spin chain. For
simplicity, we assume that the spin chain subsystem begins with its ground
state. In the following discussion, we will focus on the quantum dynamics of
the LE in the different parameter regions. In particular, the decay problem of
LE induced by the coupling of the central spin and its surrounding spin chain,
as has been discussed in Ref.[15] for the special case of Ising model, will be
fully studied in the ($\gamma,\lambda$)-space.

To diagonalize the effective Hamiltonians $H_{i}$ ($i=g,e$), we follow the
standard procedure\cite{Sach} by defining the conventional Jordan-Wigner (JW)
transformation
\begin{equation}
\sigma_{l}^{x}=\underset{m<l}{\prod}(1-2a_{m}^{+}a_{m})\left(  a_{l}+a_{l}%
^{+}\right)  , \tag{4a}%
\end{equation}%
\begin{equation}
\sigma_{l}^{y}=-i\underset{m<l}{\prod}(1-2a_{m}^{+}a_{m})\left(  a_{l}%
-a_{l}^{+}\right)  , \tag{4b}%
\end{equation}%
\begin{equation}
\sigma_{l}^{z}=1-2a_{l}^{+}a_{l}. \tag{4c}%
\end{equation}
which maps spins to one-dimensional spinless fermions with creation
(annihilation) operators $a_{l}^{+}$ ($a_{l}$). After a straightforward
derivation, the effective Hamiltonians read%
\begin{align}
H_{i}  &  =-\overset{N}{\underset{l=1}{\sum}}[(a_{l+1}^{+}a_{l}+a_{l}%
^{+}a_{l+1})+\gamma(a_{l+1}a_{l}+a_{l}^{+}a_{l+1}^{+})\tag{5}\\
&  +\left(  \lambda+\kappa_{i}\delta\right)  (1-2a_{l}^{+}a_{l})]-\kappa
_{i}\Delta,\nonumber
\end{align}
where $\kappa_{g}=-\kappa_{e}=1$. Next we introduce Fourier transforms of the
fermionic operators described by $d_{k}=\frac{1}{\sqrt{N}}\sum_{l}%
a_{l}e^{-i2\pi lk/N}$ with $k=-M,..,M$, $M=N/2$. The Hamitionians (4) can be
diagonalized by transforming the fermion operators in momentum space and then
using the Bogoliubov transformation. The results are
\begin{equation}
H_{i}=\sum_{k}2\Lambda_{k,i}(b_{k,i}^{+}b_{k,i}-1/2)-\kappa_{i}\Delta, \tag{6}%
\end{equation}
where the energy spectrums $\Lambda_{k,i}$ ($i=g,e$) are given by%
\begin{equation}
\Lambda_{k,i}=\sqrt{\epsilon_{k,i}^{2}+\gamma^{2}\sin^{2}\frac{2\pi k}{N}%
}\text{ with }\epsilon_{k,i}=\lambda-\cos\frac{2\pi k}{N}+\kappa_{i}\delta,
\tag{7}%
\end{equation}
and the corresponding Bogoliubov-transformed fermion operators are defined by
\begin{equation}
b_{k,i}=\cos\frac{\theta_{k}^{(i)}}{2}d_{k}-i\sin\frac{\theta_{k}^{(i)}}%
{2}d_{-k}^{+} \tag{8}%
\end{equation}
with angles $\theta_{k}^{(i)}$ satisfying $\cos\theta_{k}^{(i)}=\epsilon
_{k,i}/\Lambda_{k,i}$. It is straightforward to see that the two sets of
normal modes are related by the equation $b_{k,e}=(\cos\alpha_{k}%
)b_{k,g}-i(\sin\alpha_{k})b_{-k,g}^{+}$ where $\alpha_{k}=(\theta_{k}%
^{(e)}-\theta_{k}^{(g)})/2$.

The ground state $|G\rangle$ of $H_{i}$ is the vacuum of the fermionic modes
described by $b_{k,i}|G\rangle_{i}=0$, and can be written as $|G\rangle_{i}=%
{\textstyle\prod\nolimits_{k=1}^{M}}
\left(  \cos\frac{\theta_{k}^{(i)}}{2}|0\rangle_{k}|0\rangle_{-k}+i\sin
\frac{\theta_{k}^{(i)}}{2}|1\rangle_{k}|1\rangle_{-k}\right)  $, where
$|0\rangle_{k}$ and $|1\rangle_{k}$ denote the vacuum and single excitation of
the $k$th mode, $d_{k}$, respectively. Note that the ground state is a tensor
product of states, each lying in the two-dimensional Hilbert space spanned by
$|0\rangle_{k}|0\rangle_{-k}$ and $|1\rangle_{k}|1\rangle_{-k}$. From the
relationship between the two Bogoliubov modes $b_{k,e}$ and $b_{k,g}$, one can
see that the ground state $|G\rangle_{g}$ of the effective Hamiltonian $H_{g}$
can be obtained from the ground state $|G\rangle_{e}$ of $H_{e}$ by the
transformation $|G\rangle_{g}=%
{\textstyle\prod\nolimits_{k=1}^{M}}
(\cos\alpha_{k}+i\sin\alpha_{k}b_{k,e}^{+}b_{-k,e}^{+})|G\rangle_{e}$.

Now we suppose that the $XY$ spin chain is initially in the ground state of
$H_{g}$, i.e., $|\varphi(0)\rangle=|G\rangle_{g}$. Then our present task is to
derive the explicit expression for LE. First one notices that the LE in Eq.
(\ref{e3}) can be rewritten as
\begin{align}
L(t)  &  =|\langle\varphi_{g}(t)|\varphi_{e}(t)\rangle|^{2}=|_{g}\langle
G|e^{-iH_{e}t}|G\rangle_{g}|^{2}\tag{9}\\
&  =|_{e}\langle G|%
{\textstyle\prod\nolimits_{k}}
(\cos\alpha_{k}-i\sin\alpha_{k}b_{-k,e}b_{k,e})e^{-iH_{e}t}%
{\textstyle\prod\nolimits_{k}}
(\cos\alpha_{k}+i\sin\alpha_{k}b_{k,e}^{+}b_{-k,e}^{+})|G\rangle_{e}%
|^{2},\nonumber
\end{align}
where the dynamical phase in $|\varphi_{g}(t)\rangle$ contributed by the time
evolution operator $e^{-iH_{g}t}$ has been eliminated by the arithmetic module
operation in $L(t)$. By using the identity $e^{-iH_{e}t}b_{k,e}^{+}e^{iH_{e}%
t}=b_{k,e}^{+}e^{-i2\Lambda_{k,e}t}$ and after a straightforward derivation,
one obtains the expression for $L(t)$ as follows%
\begin{align}
L(t)  &  =|_{e}\langle G|%
{\textstyle\prod\nolimits_{k}}
(\cos\alpha_{k}-i\sin\alpha_{k}b_{-k,e}b_{k,e})(\cos\alpha_{k}+ie^{-i4\Lambda
_{k,e}t}\sin\alpha_{k}b_{k,e}^{+}b_{-k,e}^{+})|G\rangle_{e}|^{2}%
\tag{10}\label{e10}\\
&  =|%
{\textstyle\prod\nolimits_{k}}
(\cos^{2}\alpha_{k}+\sin^{2}\alpha_{k}e^{-i4\Lambda_{k,e}t})|^{2}\nonumber\\
&  =%
{\textstyle\prod\nolimits_{k=1}^{M}}
\left[  1-\sin^{2}\left(  2\alpha_{k}\right)  \sin^{2}\left(  2\Lambda
_{k,e}t\right)  \right]  .\nonumber
\end{align}
Remarkably, the expression for $L(t)$ based on $XY$ spin chain is formally
same as that based on Ising model which has been previously
reported\cite{Quan}. The difference comes from the time-dependent phase
factor, which in the present case is the energy spectrum $2\Lambda_{k,e}$ of
$XY$ spin-chain characterized by the effective Hamiltonian $H_{e}$, instead of
Ising model given in Ref.\cite{Quan}. Due to the obvious difference in the
energy spectrum between $XY$ model and Ising model, one may expect that the
behavior of the LE in the present case will include new features
characteristic of the $XY$ model.

Since each factor $F_{k}$ in Eq. (\ref{e10}) has a norm less than unity, we
may expect $L(t)$ to decrease to zero in the large $N$ limit under some
reasonable conditions. This kind of factorized structure was first discovered
and systematically studied\cite{Sun} in developing the quantum measurement
theory in classical or macroscopic limit and has been applied to analyze the
universality of decoherence influence from environment on quantum
computing\cite{Sun2}. Now we study in detail the critical behavior of of LE
near the critical point $\lambda_{c}=1$ for finite lattice size $N$ of spin
chain. Following Ref.\cite{Quan}, let us first make a heuristic analysis of
the features of the LE. For a cutoff frequency $K_{c}$ we define the partial
product for the LE%
\begin{equation}
L_{c}(t)=%
{\textstyle\prod\nolimits_{k=1}^{K_{c}}}
F_{k}\geq L_{c}(t), \tag{11}%
\end{equation}
and the corresponding partial sum $S(t)=\ln L_{c}\equiv-%
{\textstyle\sum\nolimits_{k=1}^{K_{c}}}
|\ln F_{k}|$. For small $k$ one has
\begin{equation}
\Lambda_{k,e}\approx|\lambda-1-\delta|+O(k^{2}), \tag{12}%
\end{equation}
and
\begin{equation}
\sin^{2}\left(  2\alpha_{k}\right)  \approx\frac{4\pi^{2}4\gamma^{2}\delta
^{2}k^{2}}{N^{2}(\lambda-\delta-1)^{2}(\lambda+\delta-1)^{2}}. \tag{13}%
\end{equation}
As a result, if $K_{c}$ is small enough one has
\begin{equation}
S(t)=-\frac{4E(N_{c})\gamma^{2}\delta^{2}\sin^{2}\left(  2t|\lambda
-\delta-1|\right)  }{(\lambda-\delta-1)^{2}(\lambda+\delta-1)^{2}}, \tag{14}%
\end{equation}
where $E(N_{c})=4\pi^{2}N_{c}(N_{c}+1)(2N_{c}+1)/(6N^{2})$. In this case, it
then follows that for a fixed $t$,
\begin{equation}
L_{c}(t)\approx\exp(-\tau t^{2}) \tag{15}\label{e15}%
\end{equation}
when $\lambda\rightarrow\lambda_{c}=1$, where $\tau=16E(N_{c})\gamma^{2}%
\delta^{2}/(\lambda+\delta-1)^{2}$.

From Eq. (\ref{e15}) it may be expected that when $N$ is large enough and
$\lambda$ is adjusted to the vicinity of the critical point $\lambda_{c}=1$,
the LE will exceptionally vanish with time. In the thermodynamic limit, i.e.,
the number $N$ of sites approaching infinite while the length of spin chain
keeping a constant, $\tau$ seems to tend to zero and thus the approximate
expression $L_{c}(t)$ remains unity without any decay. This implies that our
heuristic analysis cannot apply to the case of thermodynamic limit, in which
case the small-$k$ approximation becomes invalid. Thus to reveal the close
relationship between the decaying behavior of LE and QPT which occur only in
the thermodynamic limit, all $k$-components of $F_{k}$ in Eq. (11) should be
included. On the other side, for a practical system used to demonstrate the
QPT-induced decay of the LE, the particle number $N$ is large, but finite, and
then the practical $\tau$ in Eq. (\ref{e15}) does not vanish.%

\begin{figure}[tbp]
\begin{center}
\includegraphics[width=0.70\linewidth]{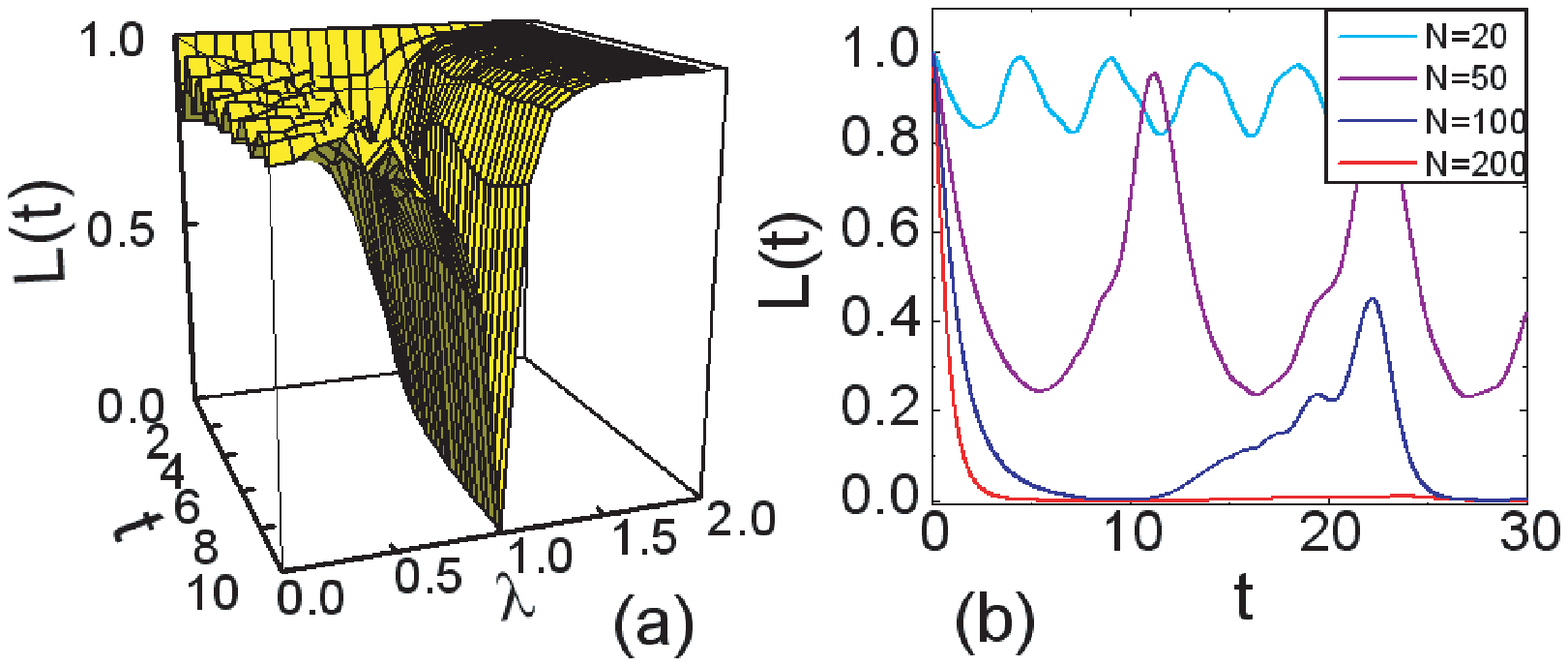}
\end{center}
\caption
{(color online). (a) The LE as a function of magnetic intensity $\lambda
$ and time $t$ for Ising ($\gamma=1.0$) spin-chain size
$N=100$; (b) The LE as a function of time for different values of N for Ising spin chain.
}
\label{fig1}
\end{figure}%
Figure 1(a) shows the numerical result of the LE in Eq. (\ref{e10}) as a
function of magnetic intensity $\lambda$ and time $t$ for $N=100$,
$\delta=0.05$, and $\gamma=1.0$ (i.e., the case of Ising model). One can see
that when the value of $\lambda$ is larger or smaller than that of
$\lambda_{c}$, the LE in time domain is characterized by an oscillatory
localization behavior. When the amplitude of $\lambda$ approaches to
$\lambda_{c}$, then the degree of localization of $L(t)$ is decreased to zero.
The fundamental change occurs at critical point of QPT, i.e., $\lambda
=\lambda_{c}=1$. At this point, as revealed in Fig. 1(a), the LE evolves from
unity to zero in a very short time. Figure 1(b) shows the time evolution of LE
for different values of lattice size at critical point $\lambda=1$ of Ising
model. One can see that the LE decays more rapidly by increasing the size $N$
of spin chain. Also the decaying amplitude is increased with increasing $N$.%

\begin{figure}[tbp]
\begin{center}
\includegraphics[width=0.70\linewidth]{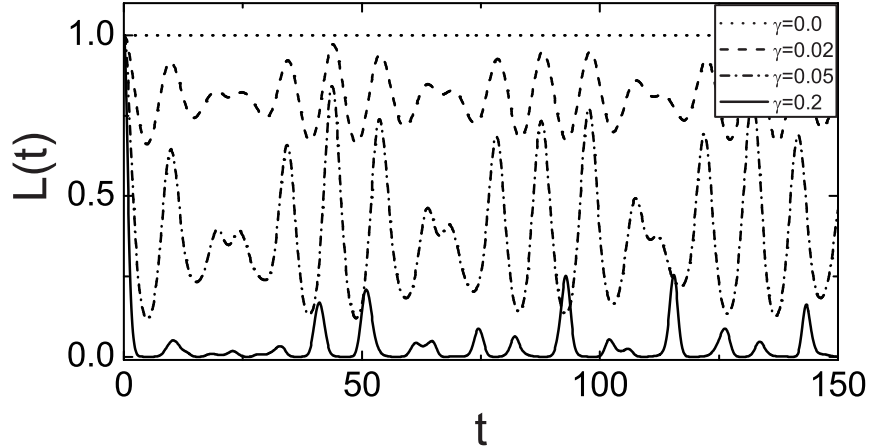}
\end{center}
\caption{The LE as a function of time for $\lambda
=1.0$ and different values of anisotropy $\gamma$.
The other parameters are chosen to be $N=100$, $\delta=0.05$.
}
\label{fig2}
\end{figure}%
Figure 2 shows the LE as a function of time for different values of anisotropy
parameter $\gamma$ in the quantum critical region ($\lambda=\lambda_{c}$). In
the extreme anisotropy limit, i.e., for the $XX$ spin model ($\gamma=0$), one
can see from Fig. 2 that the LE completely remains to be unity during time
evolution. This full localization behavior can also be seen from the analytic
expression Eq.(14), in which $\tau=0$ for $\gamma=0$, indicating no decay in
the LE, regardless of the variation of $\lambda$ and the size of spin chain.
As a consequence, the purity $P$ of the central spin remains unity; the
coupling induced decoherence disappears for the $XX$ spin chain. In this case,
the quantum criticality behavior of the surrounding spin chain dose not affect
the localization behavior of the LE for the central spin. By smoothly tuning
the value of $\gamma$ little out of $XX$ model, as shown in Fig. 2, the
behavior of the LE begins to be characterized by an interplay of the decay in
a short time and the oscillations in the subsequent evolution. The
oscillations are featured by a superposition of the collapses and the
revivals. The amplitude of the oscillations is decreased with increasing the
value of $\gamma$. Further increasing the value of $\gamma$ will, as one can
see from Fig. 2, lead to the complete decay of the LE without prominent
revivals during the whole time evolution. Therefore, the decay of the LE and
its proximity to the quantum criticality can be tuned by the anisotropy
parameter $\gamma$.

Now we turn to study the behavior of the ground-state BP for the central spin.
Due to the coupling, it is expected that the BP for the central spin will be
profoundly influenced by the occurrence of QPT in spin-chain environment.

Similar to the above discussions, it is supposed that the $XY$ spin chain is
adiabatically in the ground state $|G(\{\theta_{k}\})\rangle_{g}$ of $H_{g}$,
which is parameterized by the series $\{\theta_{k}\}$ in the ground state.
Thus the effective mean-field Hamiltonian for the central spin is given by%
\begin{align}
H_{eff}  &  =H_{S}+_{g}\langle G|H_{I}|G\rangle_{g}\tag{16}\\
&  =\left(  \frac{\mu}{2}+\frac{2g}{N}\sum_{k=1}^{M}\cos\theta_{k}%
^{(g)}\right)  \sigma^{z}+\frac{\nu}{2}\sigma^{x}.\nonumber
\end{align}
In order to generate a BP for the central spin, we change the Hamiltonian by
means of a unitary transformation:%
\begin{equation}
U(\phi)=\exp\left(  -i\frac{\phi}{2}\sigma_{z}\right)  , \tag{17}%
\end{equation}
where $\phi$ is a slowly varying parameter, changing from $0$ to $2\pi$. The
transformed Hamiltonian can be written as
\begin{align}
H_{eff}(\phi)  &  =U^{+}(\phi)H_{eff}U(\phi)\tag{18}\\
&  =\left(  \frac{\mu}{2}+\frac{2g}{N}\sum_{k=1}^{M}\cos\theta_{k}%
^{(g)}\right)  \sigma^{z}+\frac{\nu}{2}(\sigma^{x}\cos\phi-\sigma^{y}\sin
\phi).\nonumber
\end{align}
The eigen-energies of the effective Hamiltonian for the central spin are given
by
\begin{equation}
E_{e,g}=\pm\sqrt{\left(  \frac{\mu}{2}+\frac{2g}{N}\sum_{k=1}^{M}\cos
\theta_{k}^{(g)}\right)  ^{2}+\frac{\nu^{2}}{4}}. \tag{19}%
\end{equation}
The corresponding eigenstates are
\begin{equation}
|g\rangle=\binom{\sin\frac{\theta}{2}}{-\cos\frac{\theta}{2}e^{-i\phi}%
},|e\rangle=\binom{\cos\frac{\theta}{2}}{\sin\frac{\theta}{2}e^{-i\phi}},
\tag{20}%
\end{equation}
where $\sin\theta=\nu/2E_{e}$.

The acquired ground-state BP for the central spin by varying $\phi$ from zero
to $2\pi$ is given by%
\begin{align}
\beta_{g}  &  =i\int_{0}^{2\pi}\langle g|\frac{\partial}{\partial\phi
}|g\rangle=\pi(1+\cos\theta)\tag{21}\\
&  =\pi\left(  1+\frac{\mu+4gf(\lambda,\gamma,N)}{\sqrt{[\mu+4gf(\lambda
,N)]^{2}+\nu^{2}}}\right)  ,\nonumber
\end{align}
where we have defined $f(\lambda,\gamma,N)=\frac{1}{N}\sum_{k=1}^{M}\cos
\theta_{k}^{(g)}$. In the thermodynamic limit, $N\rightarrow\infty$, the
summation in $f(\lambda,\gamma,N)$ can be replaced by the integral as follows:%
\begin{equation}
f(\lambda,\gamma,N)|_{N\rightarrow\infty}=\frac{1}{2\pi}\int_{0}^{\pi}%
\frac{\lambda-\cos\varphi}{\sqrt{(\lambda-\cos\varphi)^{2}+\gamma^{2}\sin
^{2}\varphi}}d\varphi. \tag{22}%
\end{equation}
The BP $\beta_{g}$ for the central spin is closely related with QPT of its
coupled spin-chain subsystem. To manifest this, we plot in Fig. 3 the BP
$\beta_{g}$ and its derivative $d\beta_{g}/d\lambda$ with respect to the field
strength $\lambda$ as a function of spin-chain parameters $\lambda$ and
$\gamma$.%
\begin{figure}[tbp]
\begin{center}
\includegraphics[width=0.70\linewidth]{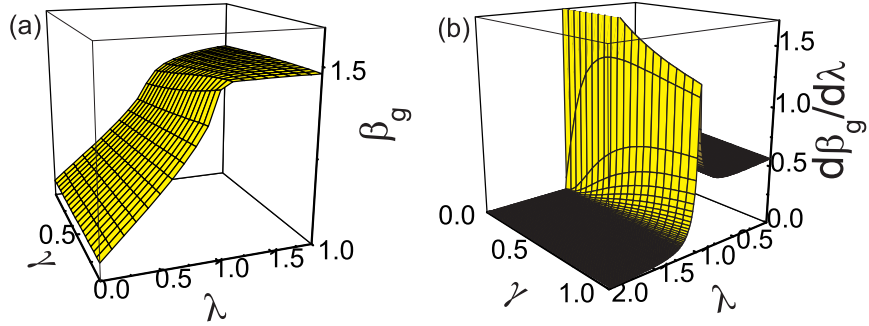}
\end{center}
\caption{(color online). (a) Ground-state BP of the central spin and (b)
its $\lambda$-derivative as a function of spin-chain parameters $\lambda
$ and $\gamma$ in the thermodynamic limit. The other parameters are chosen
to be $\mu=0.1$, $\nu=2.0$, and $g=0.5$.
}
\label{fig3}
\end{figure}
One can see that given the value of $\gamma$, the BP of the central spin
increases with increasing the field strength $\lambda$. After passing through
the critical line $\lambda_{c}=1$, the BP $\beta_{g}$ arrives at a stable
value which turns out to be determined by the specific values of central-spin
parameters $\mu$ and $\nu$. The nonanalytic property of BP and its $\lambda
$-derivative along the whole critical line can be clearly seen from Fig. 3.
Thus a nonanalytic ground-state GP $\beta_{g}$ and the corresponding anomaly
in its derivative $d\beta_{g}/d\lambda$ for the central spin also witness QPT
of the coupling spin-chain subsystem.

To help further illustration, let us consider the most discontinuous case of
$XX$ spin model ($\gamma=0$). In the thermodynamic limit, the function $f$
[Eq. (22)] occurred in the expression of $\beta_{g}$ can be obtained
explicitly for $\gamma=0$ as $f=1/2-\arccos(\lambda)/\pi$ when $\lambda\leq1$
and $f=1/2$ when $\lambda>1$. Thus the BP of the central spin is given by
\begin{equation}
\beta_{g}\bigr |_{N\rightarrow\infty}=\left\{
\begin{array}
[c]{l}%
\pi\left(  1+\frac{\mu+2g[1-2\arccos(\lambda)/\pi]}{\sqrt{\left(
\mu+2g[1-2\arccos(\lambda)/\pi]\right)  ^{2}+\nu^{2}}}\right)  \text{
\ \ (}\lambda\leq1\text{)}\\
\pi\left(  1+\frac{\mu+2g}{\sqrt{\left(  \mu+2g\right)  ^{2}+\nu^{2}}}\right)
\text{ \ \ \ \ \ \ \ \ \ \ \ \ \ \ \ \ \ \ \ (}\lambda>1\text{)}%
\end{array}
\right.  , \tag{23}%
\end{equation}
which clearly shows a discontinuity at $\lambda=\lambda_{c}=1$. On the other
side, one can see that the value of function $f(\lambda,\gamma,N)$ in
$\beta_{g}$ is always trivial for $\gamma=0$ and every finite lattice size
$N$, since $\theta_{k}^{(g)}=0$ or $\pi$ for every $k$. The difference between
the finite and infinite lattice size can be understood, as has been first
demonstrated in Ref.\cite{Zhu}, from the two limits $N\rightarrow\infty$ and
$\gamma\rightarrow0$.%
\begin{figure}[tbp]
\begin{center}
\includegraphics[width=0.70\linewidth]{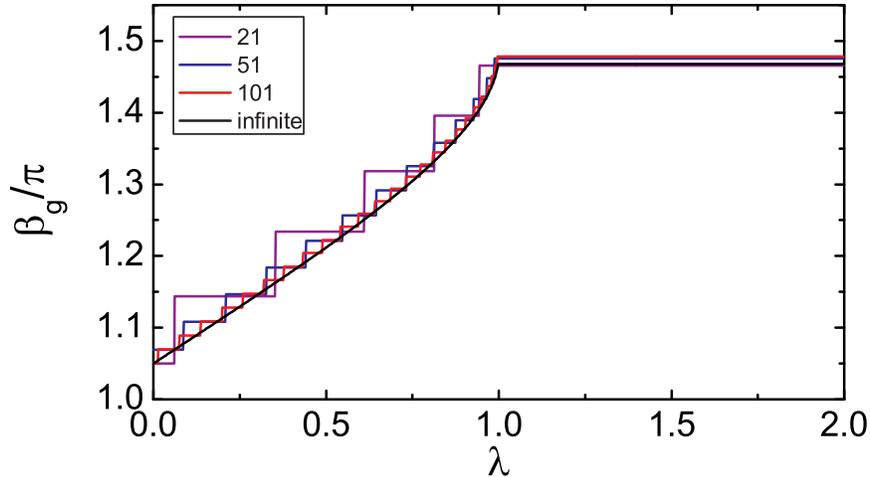}
\end{center}
\caption{(color online). $\lambda$-dependence of ground-state BP
of the central spin coupled to a $XX$ spin chain ($\gamma
=0$) with different chain sizes $N$. The
other parameters are the same as used in Fig. 3.
}
\label{fig4}
\end{figure}
We plot in Fig. 4 the numerical results of the BP $\beta_{g}$ for different
values of spin-chain size $N$, in comparison with the result for the
thermodynamic limit. One can see that the BP of the central spin displays a
multi-step like behavior for the small values of spin chain size $N$. By
increasing $N$, the BP approaches towards the case of thermodynamic limit with
nonanalyticity only at $\lambda_{c}$. We notice that the multi-step behavior
of $\beta_{g}$ for finite lattice size is a unique feature of the $XX$ model
($\gamma=0$), and will be completely washed out by deviation of $\gamma$ from zero.%

\begin{figure}[tbp]
\begin{center}
\includegraphics[width=0.70\linewidth]{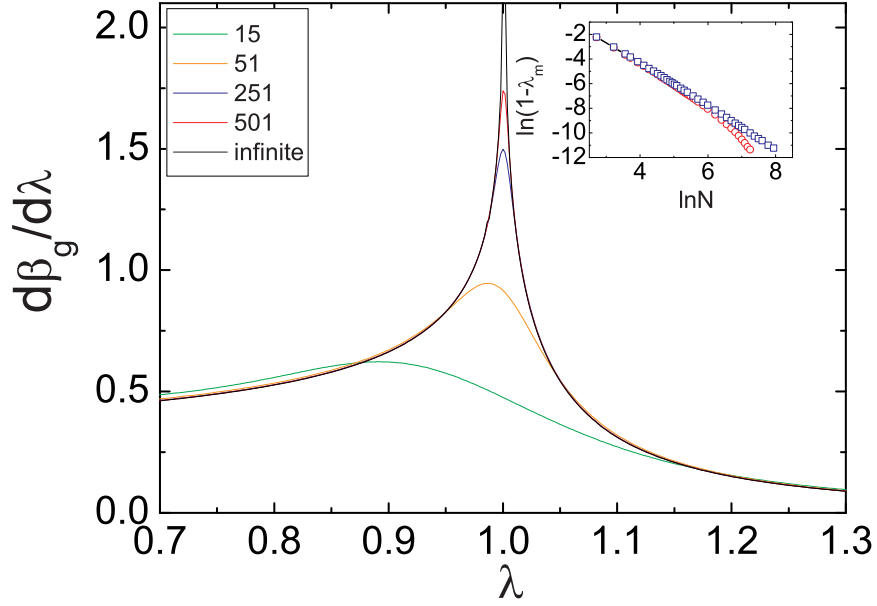}
\end{center}
\caption{(color online). $\lambda$-dependence of $d\beta_{g}/d\lambda$
for the central spin which is coupled to the Ising spin chain
($\gamma=1$) with different chain sizes $N=15,51,251,501,\infty$.
The behavior of $d\beta_{g}/d\lambda
$ for the central spin reflects QPT of spin chain. With increasing the chain sizes, the peak becomes
more pronounced. The inset shows the size scaling of the position of the peak
occurred in $d\beta_{g}/d\lambda$ (circles) and in function $f(\lambda
,\gamma,N)$ (squares).
}
\label{fig5}
\end{figure}%
To further understand the relationship between BP of the central spin and
quantum criticality of the coupled spin chain, we calculate the derivative
$d\beta_{g}/d\lambda$ as a function of $\lambda$ for $\gamma=1$ (Ising model)
and different lattice sizes. The results are plotted in Fig. 5. Two prominent
features can be seen: (i) The derivative $d\beta_{g}/d\lambda$ of GP is peaked
around $\lambda=1$, as in the thermodynamic limit shown in Fig. 3(b). The
amplitude of the peak is prominently enhanced by increasing the lattice size
of spin chain; (ii) The accurate position $\lambda_{m}$ of the peak in
$d\beta_{g}/d\lambda$ is changed with changing the size $N$ of the spin chain.
The position $\lambda_{m}$ of the peak can be regarded as a pseudocritical
point\cite{Barber}. We show in the inset (red circles) in Fig. 5 the size
dependence of the peak position $\lambda_{m}$ for $d\beta_{g}/d\lambda$. For
comparison, we also plot in this inset the size dependence of the peak
position in $\lambda$-space for the $\lambda$-derivative of quantity
$f(\lambda,\gamma,N)$. It has been shown in Ref.\cite{Zhu} that the quantity
$f(\lambda,\gamma,N)$ is proportional to the ground-state BP for the spin
chain (instead of that for the central spin discussed here) and the peak
position $\lambda_{m}$ in $df(\lambda,\gamma,N)/d\lambda$ tends as
$N^{-1.803}$ towards the critical point. This scaling behavior of
$df(\lambda,\gamma,N)/d\lambda$ is also clearly shown in the inset in Fig. 5.
Remarkably, compared to the scaling behavior of $df(\lambda,\gamma
,N)/d\lambda$, i.e., the scaling behavior of $\lambda$-derivative of
ground-state BP for spin chain, the peak position $\lambda_{m}$ in $d\beta
_{g}/d\lambda$ in the present case approaches the critical point $\lambda_{c}$
more rapidly, which is verified by the fact that in the inset in Fig. 5 the
quantity $\log(1-\lambda_{m})$ characterizing the scaling of $d\beta
_{g}/d\lambda$ curves down more rapidly than that characteristic of
$df(\lambda,\gamma,N)/d\lambda$ at large values of spin chain size $N$. Thus
we can see that QPT of the XY spin chain is reflected faithfully by the
behavior of the ground-state BP and its $\lambda$-derivative of the coupled
central spin.

In summary, we have analyzed the behavior of the Loschmidt echo in a coupled
system consisting of a central spin and its surrounding environment
characterized by a general $XY$ spin chain. The exact expression of the LE has
been obtained. The relation between the behavior of the LE and the occurrence
of QPT in spin chain has been illustrated. The decay of LE, which is closely
associated with the entanglement between the two coupled subsystems, has been
shown to be monotonically modulated by the anisotropic parameter $\gamma$ of
the spin chain. At $\gamma=0$ ($XX$ model), in particular, both the heuristic
analysis and the numerical calculation show that the LE is completely
localized to be unity without any decay. Furthermore, we have investigated the
behavior of the ground-state BP $\beta_{g}$ of the central spin. It has been
shown that the behavior of $\beta_{g}$ and its derivative with respect to the
magnetic intensity $\lambda$ of the spin chain has a direct connection with
QPT of the spin-chain subsystem. This connection is verified by the common
feature that both BP (and its $\lambda$-derivative) of the central spin and
QPT of the coupling spin chain is characterized by nonanalytic behavior around
the critical point (or critical line) $\lambda=\lambda_{c}$. Thus the QPT of
the spin chain can be revealed by studying the BP behavior of the coupled
central spin.

This work was supported by CNSF No. 10544004 and 10604010.\

\end{document}